\documentclass[]{article}
\usepackage{graphicx}
\begin{document}

\title{Causation and Physics}
\author{\it Cynthia K. W. Ma\thanks{Department of Philosophy, Logic and Scientific Method and Center
for Philosophy of Natural and Social Science, London School of
Economics and Political Science, London WC2A 2AE, United Kingdom.
\newline \hspace*{5mm} E-mail: C.K.MA@LSE.AC.UK}}
\date{}\maketitle

\begin{center}
{\bf Abstract}
\end{center}

\vspace{1mm}
\begin{center}
\parbox{14cm}{\hspace*{5.5mm}\small Philosophical analyses of
causation take many forms but one major difficulty they all aim to
address is that of the spatio-temporal continuity between causes
and their effects. Bertrand Russell in 1913 brought the problem to
its most transparent form and made it a case against the notion of
causation in physics. In this essay, I focus on this subject of
causal continuity and its related issues in classical and quantum
physics.}
\end{center}

\bigskip \bigskip

\hspace*{-5.5mm} I.  {\bf Introduction}

\hspace*{-5.5mm} II. {\bf Causal Connotations and David Hume's
Theory of Causation}

\hspace*{-5.5mm} III. {\bf Russell's Objection to Hume's Temporal
Contiguity Thesis}

\hspace*{-5.5mm} IV.  {\bf Causal Continuity and Recent
Physicalist Accounts of Causation}

\hspace*{-5.5mm} V. {\bf Concluding Remarks}

\bigskip \bigskip

\hspace*{-5.5mm}{\bf Acknowledgments}

\bigskip

This essay is inspired by the kind invitation of the organizers,
especially Professor Francesco de Martini, to speak at the III
Adriatico Research Conference on Quantum Interferometry. I am
indebted to my thesis advisor, Professor Nancy Cartwright for
introducing me to this research and her constant inspiration. I am
also thankful for the useful comments and encouragement received
from Professors Miles Blencowe, Carl Hoefer, Drs. Joseph
Berkovitz, Etienne Hofstetter and Makoto Itoh. Professor Yanhua
Shih has suggested a number of important modifications for which I
am very grateful, and for an important point brought to my
attention by Professor Asher Peres during the discussion session
that has opened new scope for my research, I extend to him my warm
gratitude.

\vspace{10mm}
\hspace*{-5.5mm}{\bf I. Introduction}
\bigskip

Causation is an active area of philosophical research and it is
one notion that is deeply entwined with the foundational aspects
of both classical and quantum physics. One need not look far but
only at the contributions to this volume that causation and its
related concepts abound in physics today.

When asked, ``What is Causation? What do we mean when we think
that one object is the {\em cause} of another or one event causes
another to happen?", no doubt different connotations would
immediately come to mind. And indeed we ought to ask ``What are
the connotations of causation?" These basic connotations, being
largely empirical in character in the sense that they come from
our experience and interactions with the physical world, generally
receive different treatments by physicists and philosophers. The
distinction is a matter of  the difference in practice. Physicists
accept these basic intuitions as facts about causation and
physical theories are constructed to conform to these ``conditions
of causality", which are not to be violated. A good example would
be that of the ``past-future" directed Minkowski light-cone
structure defined in terms of ``cause-effect" relations\footnote
{See for example, {\em Taylor and Wheeler} (1966), Spacetime
Physics, p.39.}. Philosophers, on the other hand, approach the
subject from a different angle; they conduct conceptual analyses
of these causal connotations and see whether they do make good
logical sense or are infected with grave inconsistencies. With the
advent of relativity theories, physics has added important items
to the stock of causal facts. Perhaps the most significant one is
that special relativity places a limit on the velocity of
propagation of causal influences. For the serious philosophical
minds, these results cannot afford to go unnoticed and it would
indeed be of considerable interest to investigate the extent of
the possible interplay between the findings from the respective
disciplines.

With this motivation in mind, the plan of this essay is as
follows. In part II, I shall consider a number of basic causal
connotations and the various aspects that may be deduced from
these considerations. We are then be placed in a position which
leads naturally to a presentation of the main ideas of  David
Hume's theory of causation. Hume's theory is the start of the
empiricist philosophical analysis that aims to capture our causal
intuitions. The Humean account and its more modern variants
collectively still represent the predominant philosophical view on
causation. However, in a cleverly written paper in 1913, Bertrand
Russell was able to show that the Humean view is not entirely free
from inconsistencies given an important assumption on the nature
of time. Russell's argument will be examined carefully in Part
III. Part IV focuses on the issue of causal continuity and the
recent physicalist approaches to causation which attempt to
resolve some of the more pressing difficulties associated with
this issue. This will then be followed by a brief conclusion of
the main points discussed and some speculative remarks in Part V.

To provide an overview of  causation is well beyond the scope of
this work but it is however my modest aim to bring into focus, in
the following pages, some major philosophical worries on the
subject which may be fairly regarded as one of the underlying
puzzles encountered in the foundations  of both classical and
quantum physics.

\bigskip \bigskip
\hspace*{-5.5mm}{\bf II. Causal Connotations and David Hume's
Theory of Causation}
\bigskip

When one event (or something\footnote {What kind of entities does
the causal relation relate is an important aspect of philosophical
analyses of causation. Some argue that the ``relata" should be
events, while others insist that they may be facts, processes or
states of affairs. However, as both Hume and Russell take events
as the proper causal relata, we may therefore consider events in
the present work.}) is regarded as the {\em cause} of another, we
have ``postulated"\footnote {Notice I have deliberately used the
word ``postulated" because it is only correct to remain
philosophically neutral and avoid making undue assumptions from
the outset.} the existence of a special relation or {\em
connection} between the two events. How special is it? We may want
to emphasize the importance of  such a connection by the
expression that the event we have chosen to call the {\em cause}
and the one that is called the {\em effect} are {\em necessarily}
connected to each other so that {\em had the cause not happened,
the effect would not have happened either}. Put slightly
differently, this counterfactual mode of representation of the
special connection refers to an element of necessity in the sense
that given the cause, the effect ``must" follow and any other
situations just simply cannot and would not happen.

How then are we to discover this necessary connection, whatever it
may be? One useful place to look is to start from our observations
of how causes and effects behave generally. An obvious observation
is that ``{\em causes precede their effects}";  namely, causes
occur earlier in time than their effects. One realises of course
that not every pair of events happening at the same two respective
instants of time are to be thought of as causes and effects. A
concrete everyday example illustrates (Fig.1). Suppose we have two
people, Angelo and Bianca, standing side-by-side in a room and
Angelo who is nearer to the light switch turns it on and at the
{\em same moment in time}, Bianca starts to sing. Even though both
Angelo's action and Bianca's singing are {\em both} events
happening {\em prior to} the lamp being lit up, we would deem it
appropriate to attribute the cause of the lamp lighting to the
switching action provided by Angelo but not to Bianca's singing.
Why? It is in part because a {\em continuous} physical connection
is envisaged between the light switch and the lamp and there is in
general no obvious and direct correlation between the processes of
singing and the lamp lighting.

So temporal succession between two events alone is not a
sufficient condition for causation. To place the matter in a more
scientific perspective, consider the Minkowski lightcone in Fig.2.
Events $E_j$  and $E_k$\footnote {These are treated as
simultaneous events - lying on the same hyperplane - but the
argument would remain valid even if they do not.} both lie in the
future of the event $E_i$ and hence are {\em causally connectible}
to $E_i$ since signals sent from  $E_j$ can reach either $E_j$ or
$E_k$. However, neither $E_j$ nor $E_j$ is {\em necessarily
connected} to $E_i$  for there {\em need not} be an existing
connection after all. Whether there is in fact an actual
connection depends upon the existence of actual physical processes
linking $E_i$ to $E_j$  and/or $E_k$. Therefore, {\bf temporal
succession and the spatio-temporal continuity between the cause
and effect provided by physical processes together seem necessary
of causation}.

\vspace{6mm}\hfill
\begin{minipage}[b]{.38\textwidth}
    \centering
    \includegraphics[height=40mm]{Figure1.eps}
    \makeatletter\def\@captype{figure}\makeatother
    \caption{}
\end{minipage}\hfill
\begin{minipage}[b]{.38\textwidth}
    \centering
    \includegraphics[height=40mm]{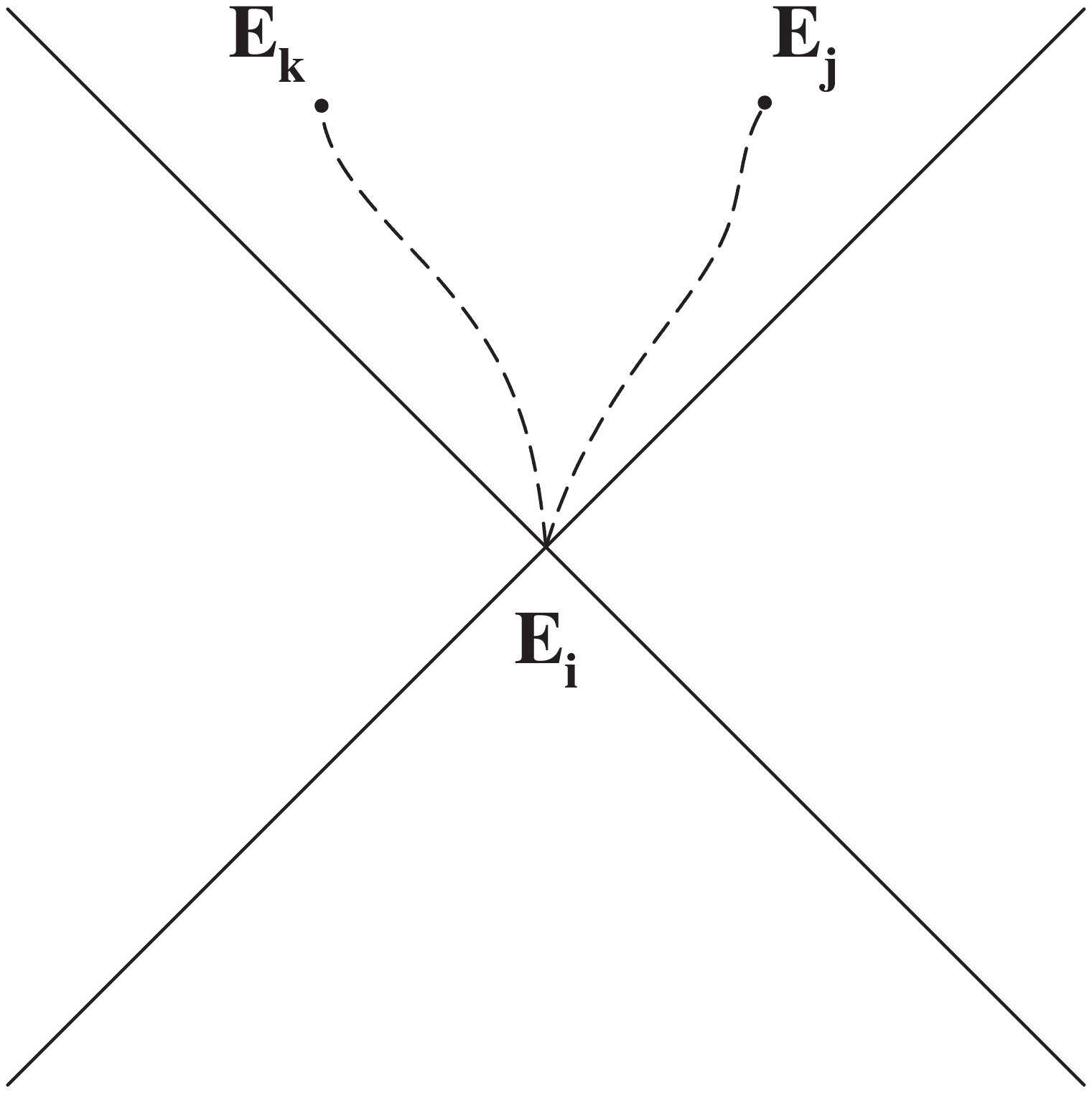}
    \makeatletter\def\@captype{figure}\makeatother
    \caption{}
\end{minipage}\hspace*{\fill}
\vspace{6mm}

Granted the physical connection between the light switch and the
lamp, is this connection really necessary, in the sense that
contrary situations are precluded from occurring? Unfortunately,
the answer is in the negative. Although it takes an awfully short
time for the electrical signal to travel from the light switch to
the lamp, it is however always conceivable that an accident like a
power cut may occur within this short time interval, as a result
of which the lamp would fail to light up. It is therefore {\em
not} necessarily the case that {\em whenever} the switch is turned
on, the lamp {\em must} light up; this is only true if no other
factors are to interfere.

Let us imagine instead the scenario where Bianca is the only
person in the room and there is no light switch attached to the
wall. We observe that when Bianca starts to sing the light comes
on a split second later. On one mere instance of this observation,
it would be reasonable to put it down as a case of sheer
coincidence because we do not usually conceive of a possible
(physical) connection between these two events. However, if such
an observation is repeated many times and in each and every time,
the same sequence obtains so that whenever Bianca starts to sing,
the light comes on, then we would conclude that the occurrences of
both events in close temporal succession are too regular to be
discounted as pure chancy coincidences. And so from repeated
observations of the regular succession of the two events, we find
it proper and indeed justified to ``infer" a special connection
(Fig.3) between this pair of events and set out to search for the
``hidden" mechanism that could have been responsible for giving
rise to such a correlation.

The question remains: is such a connection we have so inferred
upon repeated observations of regularities a {\em necessary} one?
Although experience teaches us that frequent correlations are
usually {\em prima facie} good indicators of causation, it is
however well-known that correlations {\em do not have} to imply
causation. Logic does not prohibit the apparent correlations we
see as arising from pure chance. One may well imagine the world to
be a chancy enterprise in such a way that ``it so happens" that
whenever $E_1$ occurs, $E_2$ follows later. Still, one may argue
that quite unbeknown to us, there could exist some kind of a
voice-recognition device which provides the physical connection
between Bianca's singing and the lighting up of the lamp. But what
gives us the impression and prompts us to look for this ``unknown
device" in the first place? Nothing other than the constant
conjunction of the two events of  the singing and the lamp being
lit - {\em given the same cause, the same effect follows}. Once
again, such a physical connection is by no means necessary as for
instance, a power cut may occur and tamper with the normal
functioning of the device, thus rendering the succession of the
two events unattainable.

\vspace{6mm} \hfill
\begin{minipage}[b]{.38\textwidth}
    \centering
    \includegraphics[height=40mm]{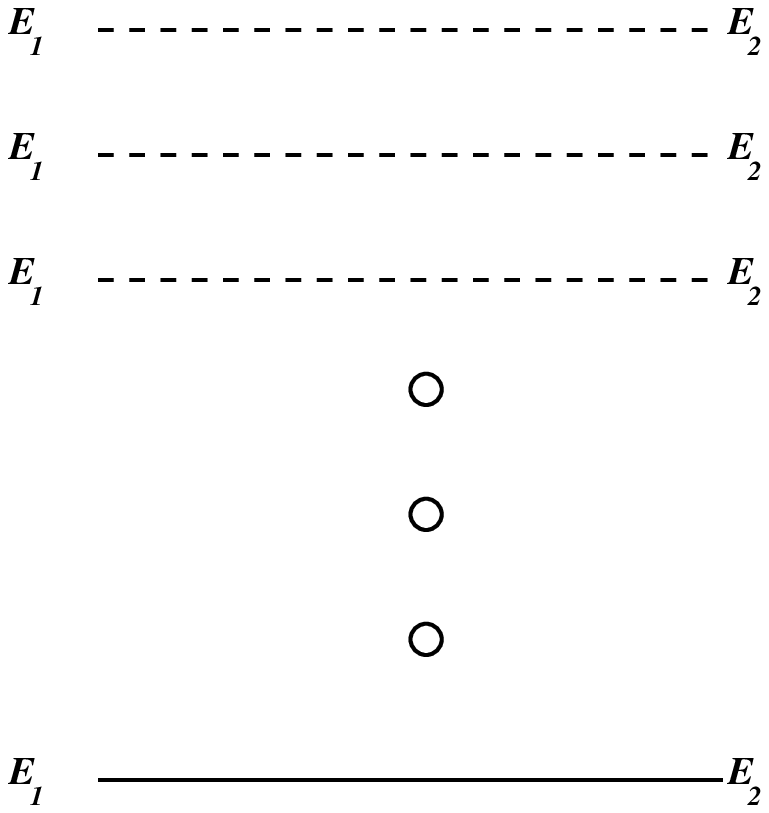}
    \makeatletter\def\@captype{figure}\makeatother
    \caption{}
\end{minipage}\hfill
\begin{minipage}[b]{.38\textwidth}
    \centering
    \includegraphics[height=40mm]{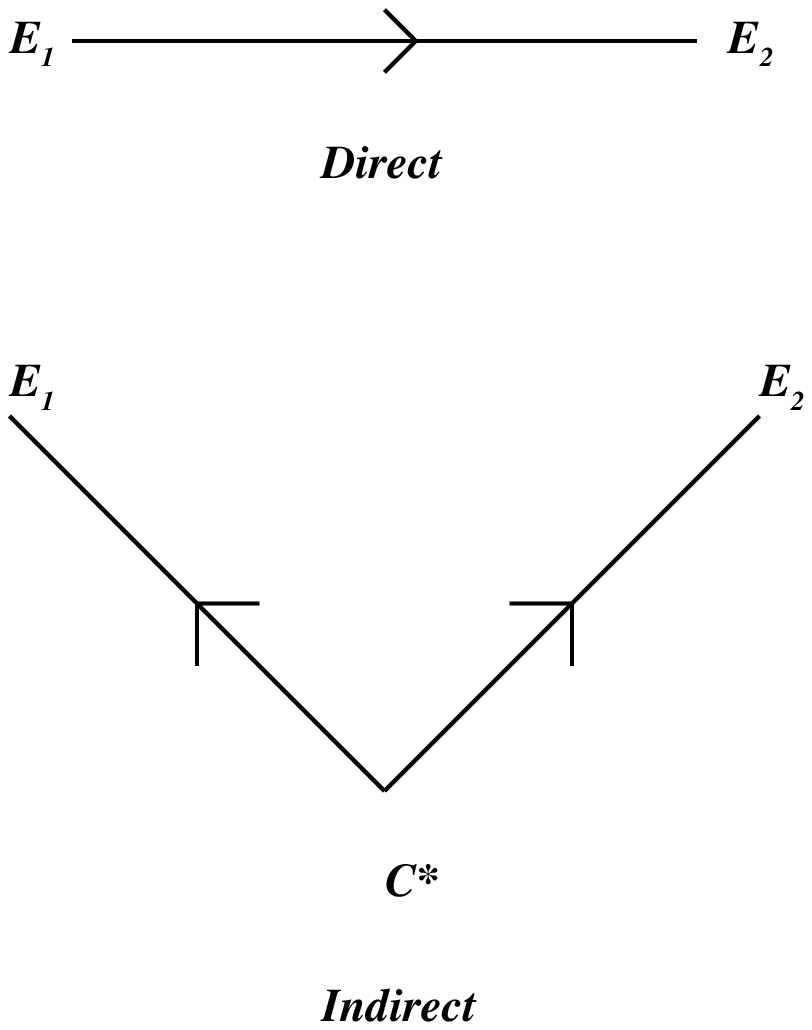}
    \makeatletter\def\@captype{figure}\makeatother
    \caption{}
\end{minipage}\hspace*{\fill}
\vspace{6mm}

Temporal succession of two events is not sufficient for causation.
A case of causation is deemed to obtain when temporal succession
is supplemented by the presence of a spatially and temporally
continuous physical connection which provides the link between the
two events in question. Similarly, the situation where the same
cause has found to be followed by the same effect on a great
number of occasions also gives us the feel of a causal correlation
and induces us to look for an underlying connection. It is in this
way that it is often thought that {\em spatio-temporal continuous
physical processes are essential for making sure that ``the same
cause is to be followed by the same effect"}. Unforeseen
circumstances may happen during the spatial and temporal course of
this continuous physical process that frustrate the connection
between the two events. Due to the absence of necessity, in the
sense that unforeseen circumstances can always occur, it follows
that there can never be any guarantee without fail that the same
cause is to be followed by the same effect. This clearly indicates
an incompatibility between these two causal connotations and that
in reality, ``the same cause is always followed by the same
effect" is not warranted by spatio-temporal continuity. This is
quite contrary to our ordinary way of thinking.

Spatially and temporally continuous physical processes, though
they are by no means necessary connections (in a logical sense),
do nevertheless impress upon us the idea of the existence of a
causal relation. Such physical connections between two events may
either be (i) direct or, (ii) indirect. In the latter case, we
seek a third event $C^*$  that existed in the overlap region of
the respective past lightcones of the two events such that both
events are direct consequences of $C^*$. In this model, both
events are effects of the common cause $C^*$ (Fig.4). This
illustrates nicely the appeal of hidden variable programs in the
efforts to provide an explanation of the Einstein-Podolsky-Rosen
(EPR) paradox. Special relativity rules out a direct physical
connection between the two space-like separated measurement
events. However, the existence of the remarkable correlations of
the results of measurements calls urgently for an explanation. It
is thus natural to look for possible common causes (hidden
variables) in the past which may have given rise to these
correlations. But, while such a procedure may satisfy one's
intellectual urge, logic permits a world in which these
correlations are all there are; the empire of chance rules in such
a manner that the correlations always obtain even without any
underlying spatio-temporal continuous connection, be it direct or
indirect. In fact, the significant achievement of Bell's 1964
inequalities rests on their success in dispelling local hidden
variable models in favour of quantum mechanics, as has since then
been so forcibly confirmed by many sophisticated experiments.

In relation to the {\em spatio-temporal continuity between causes
and effects}, another common connotation we have for causation is
to suppose that a certain cause event has occurred but
nevertheless the expected effect has somehow failed to
materialize. Then we usually explain this by saying that other
events must have got in-between them which inhibited the
occurrence of the effect. In other words, the spatio-temporal
continuity between the supposed cause and the supposed effect is
disturbed. In order to diminish the opportunity for other events
to get ``in-between" and be mischievous, we want to make both the
spatial and temporal intervals between the cause and the effect as
short as possible. The shorter these intervals are, generally the
smaller the probability for other factors to interfere. So, causes
and effects are expected to be spatially and temporally close to
each other, that is, they should exhibit a degree of
spatio-temporal ``nearness". Remote factors are not immediate
causes and it is in this way that causal continuity is to be so
intimately bound up with the notion of locality. For events
happening at vastly separated spacetime locations to be causally
connected, we look for events to provide the intermediate links
between these two spacetime locales; hence the idea of a ``causal
chain" to ensure causal continuity across spacetime regions.

The foregoing discussions now lead appropriately to the
introduction of the major ideas of David Hume's theory of
causation, which exerted tremendous influence over the logical
positivists and their contemporaries such as Bertrand Russell.

Hume maintains that there are two basic elements to human
understanding that form the pillars to his philosophical system:
{\em impressions} and {\em ideas}. Impressions correspond to all
``lively signals" we receive from the physical world through our
senses, like perceptions, sensations, feelings etc. Ideas, on the
other hand, consist in the formation of a conception of
impressions. The general principle he adopts for his philosophical
analyses is that ``{\em all ideas originate from the association
and combination of the different impressions}". That is to say, a
certain idea we may have for something has to come from our {\em
experience} through our senses.

Armed with this principle, he then asks {\em from which
impressions} do we form the idea of cause-and-effect as some sort
of a {\em necessary connection}? He is able to identify three such
impressions from our empirical experience behaving like causes and
effects. These are: ``{\bf priority in time}" of the cause ,
``{\bf constant conjunction}" between the cause and effect and
``{\bf contiguity}" in space and time between the causes and
effects. And these should be of some familiarity since they refer
to none other than the three causal connotations we have
considered in the above: {\em causes precede their effects, given
the same cause, the same effect follows} and {\em continuity}
respectively. But as we have already discussed, from these three
properties and these three alone, one can never deduce the element
of necessity. Hume argued that even though there may actually
exist connections in the world which are necessary in some sense,
beyond this the only real idea we can have of this connection is
of the three properties above. Since these properties are not
sufficient to entail necessity, philosophical prudence must now
compel us to take a skeptical view of the idea of necessary
connection between causes and effects.

Granted that our experience is incapable of furnishing us with the
idea of a necessary causal connection, how are we able to
associate the three impressions of causes and effects to arrive at
the idea of a necessary connection between two events? Hume's
answer consists in the fact that after many instances of observing
the behaviours of constant conjunction, priority in time and
contiguity in time and space between the two events {\em without
exception}, the mind has in the course grown accustomed to expect
that a special connection does indeed exist between the two. This
feeling of expectation then gives us the impression from which our
idea of connection is copied. The idea is thus not from our
experiences of the external world but comes rather from our own
response to it. In a sense then the causal relation as a necessary
connection is an idea ``imposed" by the mind upon unfailing,
successive observations of these regular behaviours of causes and
effects. The three impressions of priority in time, constant
conjunction and contiguity in time and space can never provide us
with  the idea of a necessary connection.

It must again be emphasized that it is never Hume's intention to
deny the existence of necessary connections in nature. Rather,
that if the three impressions are all we have by way of the
evidence for causal necessity, and since this evidence alone is
not adequate for us to serve to reveal to us such an element of
necessity, it would be more reasonable not to impose their
existence on nature, leaving this instead as an open question. And
Hume concludes\footnote{David Hume, (1888), {\em A Treatise of
Human Nature}, second edition (1978), with revised text and notes
by P.H. Nidditch, Clarendon Press, Oxford. (THN)},

\bigskip

\begin{center}
\parbox{14cm}{\em
``As to what may be said, that the operations of nature are
independent of our thought and reasoning, I allow it; and
accordingly have observed, that objects bear to each other the
relations of contiguity and succession; that like objects may be
observed in several instances to have like relations; and that all
this is independent of, and antecedent to the operations of the
understanding. But if we go any farther, and ascribe a power or
necessary connection to these objects; this is what we can never
observe in them, but must draw the idea from what we feel
internally in contemplating them." (THN, p.168-9)}
\end{center}

\bigskip \bigskip
\hspace*{-5.5mm}{\bf III. Russell's Objection to Hume's Temporal
Contiguity Thesis}
\bigskip

In 1912, Russell made his presidential address\footnote{Delivered
on 4 November 1912. The ensuing essay was published in the {\em
Proceedings of the Aristotelian Society}, 13 (1912-13) and
reprinted in Russell, B. (1917), {\em Mysticism and Logic},
p.180-208, George Allen and Urwin.} to the Aristotelian Society
the occasion to cast doubt on the tenability of the Humean account
of causation and to argue against the notion of cause in physics.

We have already taken pains to stress the inherent
incompatibilities among the three causal impressions of priority
in time, contiguity in space and time and constant conjunction. In
particular, it has been indicated that given the absence of the
ingredient of necessity, spatio-temporal continuity is not really
capable of  ensuring the constant conjunction of the causes and
effects. The main reason for this is that even if there is a
continuous spatio-temporal physical process connecting the cause
and the effect, anything can still happen during the time interval
when the causal influence is transmitted down the connection and
this results in an uncertainty in the production of the effect.
The light switch and the lamp in the last section form a good
example. This is why events which are too removed from each other
in both the spatial and temporal dimensions are not considered as
reliable causes and effects.

An immediate solution would be to require that both the spatial
and temporal distances between the two events be decreased to such
an extent that they stand ``{\em adjacent}" (or {\em contiguous})
to each other so that we may have the assurance that other factors
cannot impose themselves and thwart the occurrence of the effect.
But what exactly does one mean by two events being ``adjacent" to
each other when embedded in a background of spacetime continuum?
The notion of spatial contiguity between two events is easily
satisfied and in the limit it is met by the case where two events
can occupy the same location when happening at different times.
The notion of temporal contiguity is however more problematic
since given that two events occur at the same spatial location
with one after another, how are we to ensure that they are
temporally contiguous to each other? So this problem reduces to
one which concerns temporal contiguity and this is indeed the
important issue addressed by Russell in his paper.

Russell's argument begins with a statement of the temporal
contiguity thesis (TC). The properties of priority in time and
temporal contiguity between cause and effect can be summarised as
follows:

\begin{center}
\parbox{14cm}{\bf TC: ``Whenever the first event (cause) ceases
to exist, the second comes into        existence immediately
after."}
\end{center}

To place TC in the correct perspective, Russell makes the major
assumption that time is to be modeled as a mathematical continuum
(MC) and is therefore considered as a dense series. A dense series
has the distinctive feature that the notion of a ``next point"
does not make sense because between any two points there always
exists others, no matter how close these two points are to each
other. It is instructive to contrast the idea of a dense series
such as  the real number line with the discrete series of positive
integers where the notion of consecutive (or ``next") members does
take on a well-posed meaning. Having specified how the temporal
continuum is to be represented, we now consider two point events
$c$ and $e$ occurring at two respective instants of time $t_1$ and
$t_2$ ($t_1 < t_2$). Because time is a dense series, it follows
that between any two instants (points) of time, there are always
other instants (points) no matter how short we make the interval
$t_2 - t_1$. That is, there is always a temporal gap between $c$
and $e$ and hence $c$ cannot be contiguous {\em in time} to $e$.
Furthermore, this temporal gap provides ample opportunities for
other events to creep in between $c$ and $e$ and to interfere.
While these other factors may not prove harmful to the production
of $e$ at $t_2$, they may however also behave otherwise and hinder
the occurrence of $e$ (Fig.5). Under these circumstances, one
cannot be certain that the same cause is always followed by the
same effect since there can always be the chance of $e$ not
occurring {\em whenever there is to be this temporal gap between
the two events}.

In order to be rid of unsolicited factors, one must devise a means
to ensure that the temporal gap is filled. An obvious way is to
suppose the cause event as having a temporal dimension (Fig.6).
The cause is now a {\em static, unchanging} event\footnote{If a
non-static, changing event such as one composed of a causal chain
of discrete events as in Fig.5, then the problem of temporal gaps
existing in-between these events within the causal chain
remains.}, occupying the half-open interval, and is imagined to
sit there from time $t_1$ to $t_2$, filling the temporal gap and
all of a sudden, turns into $e$ at $t_2$. However, Russell objects
strongly to such kinds of events: he argues that it is not at all
logical why, being unchanging and sitting there complacently, $c$
has to turn into $e$ at $t_2$ and not at any other moments, say
earlier at $t_0$ or later at $t_3$?

\vspace{6mm} \hfill
\begin{minipage}[b]{.32\textwidth}
    \centering
    \includegraphics[width=50mm]{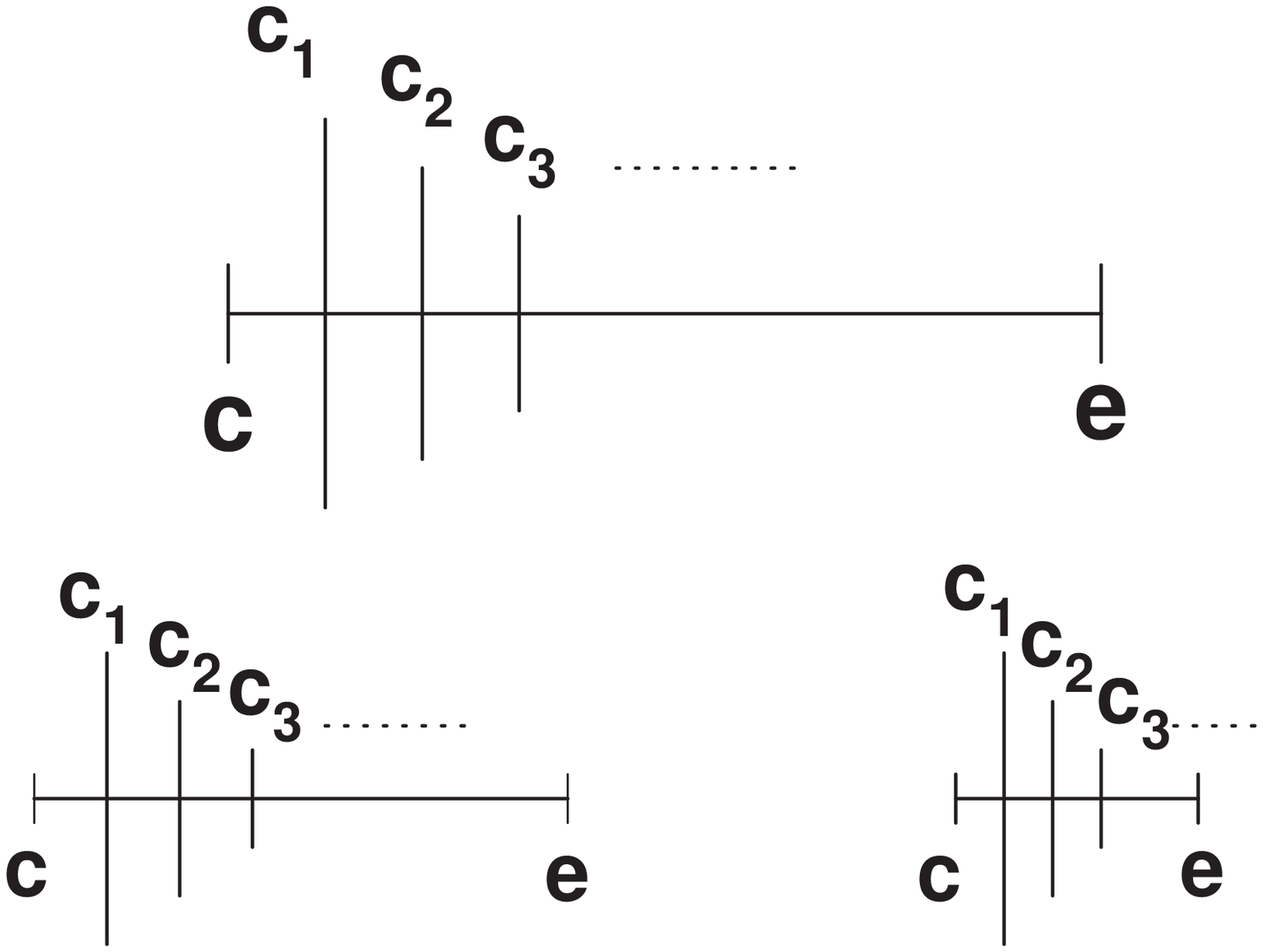}
    \makeatletter\def\@captype{figure}\makeatother
    \caption{}
\end{minipage}\hfill
\begin{minipage}[b]{.32\textwidth}
    \centering
    \includegraphics[width=50mm]{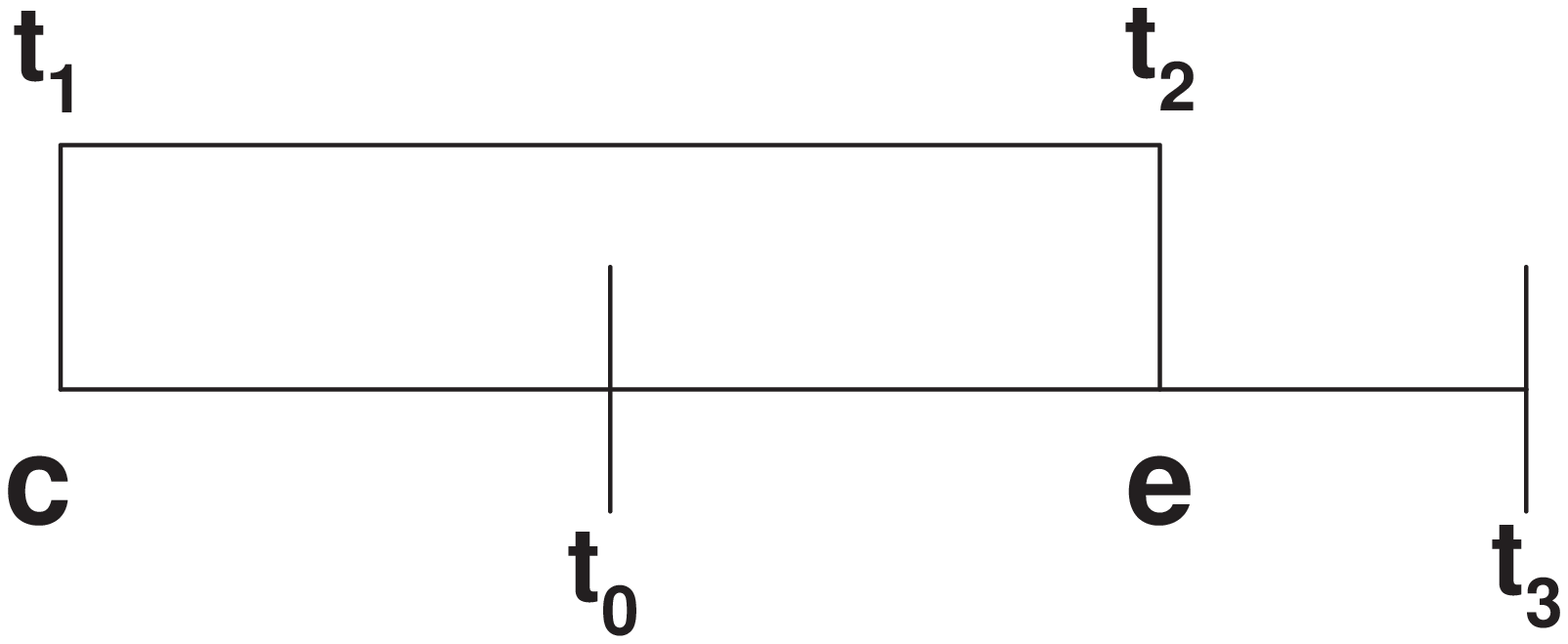}
    \makeatletter\def\@captype{figure}\makeatother
    \caption{}
\end{minipage}\hspace*{\fill}
\vspace{6mm}

And so static, unchanging events are dismissed outright by Russell
as an impossibility. Since these static, unchanging events which
seem to be the only means by which the temporal gap can be
occupied are not plausible, we must therefore draw the conclusion
that there always exists a temporal gap between $c$ and $e$ so
that $c$ cannot be contiguous to $e$. Our intuition about the
temporal continuity of causes and effects comes under threat given
the assumption of physical time as a mathematical continuum and
constant conjunction cannot be guaranteed under such a
circumstance. Russell has succeeded in showing that there exist
tremendous tensions between our usual connotations of the causal
relation.

\bigskip \bigskip
\hspace*{-5.5mm}{\bf VI. Causal Continuity and Recent Physicalist
Accounts of Causation}
\bigskip

Despite the difficulty brought to light by Russell's critique of
the Humean temporal contiguity thesis, one is, of course, allowed
to argue that the major issue is really the definition of events
as points occurring at {\em discrete} temporal instants within the
temporal continuum. There is simply no place for the notion of
discreteness with a temporal continuum. So, a more amicable
approach would be to ``superpose" a continuum of events - a
continuous rope of events - upon this temporal continuum in the
sense that we consider {\em all} the events that happen locally
within this time interval. Here we focus more closely on the
aspect of causal continuity by first considering this example from
Elizabeth Anscombe\footnote{Anscombe, E (1974), "Times, Beginnings
and Causes" in the {\em Proceedings of the British Academy}.
Reprinted in G.E.M. Anscombe, {\em Metaphysics and the Philosophy
of Mind}, Collected Philosophical Papers Volume II, p.148-162.}
(1974, p.150): ``Find an object here and ask how it comes to be
there?" (Fig.7)

\begin{figure}[hbt]
  \centering
    \includegraphics[width=60mm]{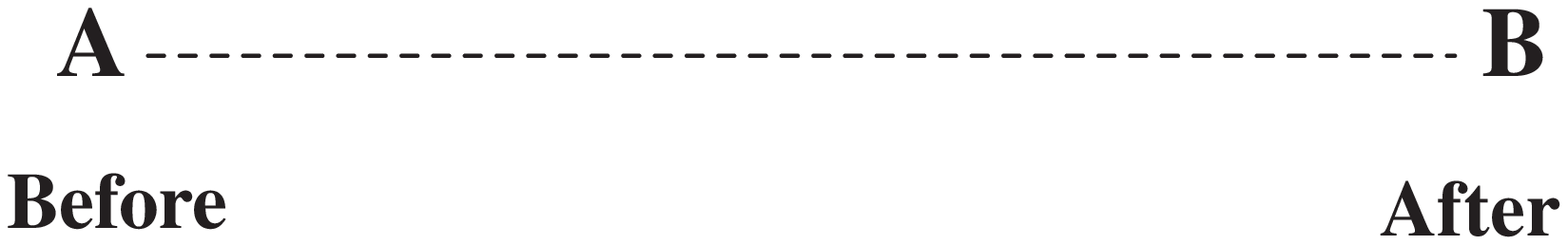}
    \caption{\label{figure7}}
\end{figure}

A {\em causal} explanation, says Anscombe, would be ``it went {\em
along some path} from {\bf A} to {\bf B}". The locution ``along
some path" in fact entails more than the case where the object
just turned up at location {\bf B} after having been at {\bf A}
previously. It requires the object to occupy also all the
intermediate positions between {\bf A} and {\bf B}. To satisfy
{\em constant conjunction}, it is sufficient for the object to
turn up at location B after having been at A some moments earlier
and {\em without} having to assume the intermediate positions
between the two locations. But this would not be deemed to be an
adequate causal explanation. And to ``explain causally", a path
has to be imposed to provide the connection between the two events
of the object being at the two respective spacetime locations. For
the purpose of explanation, it is therefore proper to consider
spatio-temporal continuous connections when thinking about
causation.

In physics, the notion of continuity is usually either represented
by spatially and temporally continuous paths or by trajectories in
phase space. These spacetime paths and phase space trajectories
are in turn the solutions of differential and integral equations.
Recalling the fact that while these differential and integral
equations guarantee continuity, they however lack the crucial
causal aspect of an explicit temporal order for cause and effect.

The pressing question which must now occupy us is how we may
introduce a temporal order into a theory of causation which takes
seriously the view that spatio- temporal continuous physical
processes, as represented by the equations of physics, provide for
us the appropriate causal connections.

We find a clue in the following consideration of the Minkowski
lightcone (Fig.8),

\begin{figure}[hbt]
\centering
    \includegraphics[height=50mm]{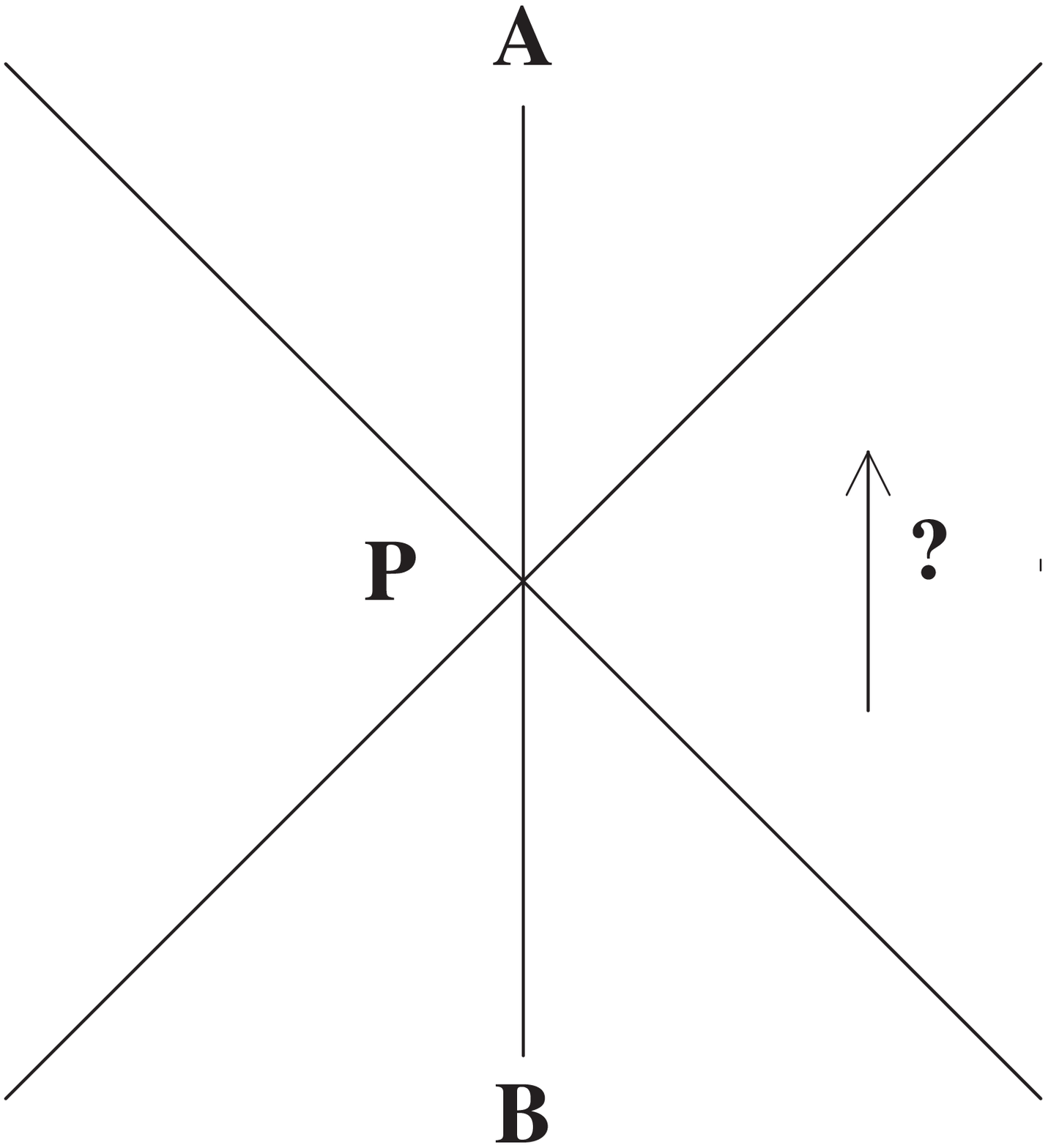}
    \caption{\label{figure8}}
\end{figure}

Does it make sense to assign a temporal direction from the past to
future as directing from the lower-half of the lightcone towards
the upper-half and not vice versa? In other words, does it make
any sense to provide the worldline as going through the
worldpoints A , B with an arrow and to assert that B is before P
and A after P? Einstein asks in 1949\footnote{Einstein A (1949),
``Reply to Criticisms" in {\em Albert Einstein:
Philosopher-Scientist}, p.687-688, The Library of Living
Philosophers Volume VII, Edited by P.A. Schilpp, Open Court.},
``Is what remains of temporal connection between world-points in
the theory of relativity an asymmetrical relation, or would one be
just as much justified, from the physical point of view, to
indicate the arrow in the opposite direction  and to assert that A
is before P, B after P?" Physics has a ready reply to this. The
``temporal arrow" is secured by the observations of irreversible
processes in nature despite the much advertised time- symmetrical
character of the laws of physics. These processes are believed to
be ultimately related to the growth of entropy in the universe.
The ``past-future" direction of the lightcone may equally well be
defined in terms of either the direction of the cause-effect
relation or that of irreversible processes. This suggests that the
causal direction may indeed be identified with the direction of
irreversible processes. Or put differently, a cause can be
considered as an event which introduces a ``change" which is
irreversible.

The idea of a cause as an irreversible change forms the backbone
of the physicalist theories of causation. Although there are
variations amongst the theories of physical causation that have
been put forward, they nevertheless share one basic underlying
idea: {\em causal continuity is guaranteed by the transmission of
causal influences (objective physical quantities) along continuous
space-time paths governed by physical laws}. The ``objective
physical quantities" being transferred refer usually to either
momentum or energy in the class of approaches subsumed under the
title of ``transference theories of causation". A slightly more
sophisticated version is the so-called ``process-theory" of Wesley
Salmon\footnote{Salmon, W C (1984), {\em Scientific Explanation
and the Causal Structure of the World}, Princeton University
Press.}; there it refers to the transmission of marks with the
marks being changes that have resulted from irreversible physical
interactions.

Interactions are responsible for bringing about or producing the
irreversible changes. To cast this concept in a better context,
consider the simple case of one mass in motion colliding with
another which is initially at rest, subsequently setting the
second into motion (Fig.9).

\begin{figure}[hbt]
    \centering
    \includegraphics[height=40mm]{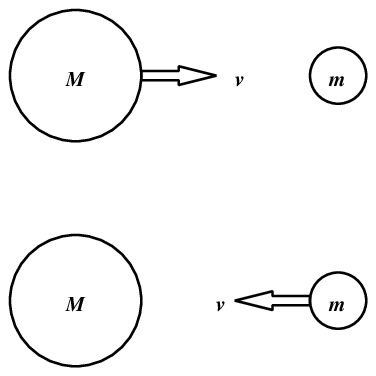}
    \caption{\label{figure9}}
\end{figure}

The same state of affairs can be described by two different causal
stories. In the rest frame of m the moving mass M travelling with
velocity {\bf v} appears to be the {\em earlier} event - the cause
which is responsible for the change of states of both masses. On
the other hand, in the rest frame of M, the {\em earlier} event of
m moving with velocity -{\bf v} is now regarded as the cause
giving rise to the subsequent change of motion of both masses.
Hence, we find ourselves confronted by two different causal
stories whose accuracies depend on the frame of reference in which
the same state of affairs is viewed. The objective matter-of-fact
is however that for all inertial frames of reference, the
``collision" between M and m produces the subsequent ``changes" of
motion of each of the masses. It is important to realise that the
{\em collision} occurs in all frames of reference and after which
is to be followed by changes in motions of  these masses. It is
indeed by this very means that an objective temporal order may
indeed be established. Because of this ``causal interaction"
between the masses M and m, their respective energies and momenta
are correlatively modified accordingly. Both masses, having
interacted, will carry the causally modified dynamical properties
via their continuous spatio-temporal trajectories and may
participate in further interactions (Fig.10).

\begin{figure}[hbt]
    \centering
    \includegraphics[height=30mm]{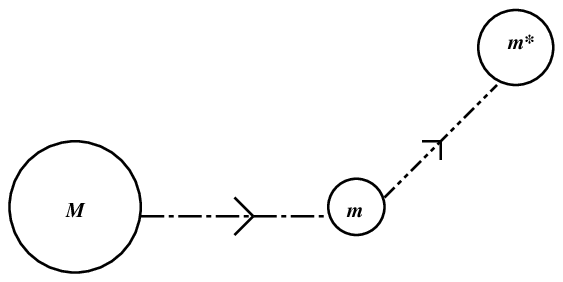}
    \caption{\label{figure10}}
\end{figure}

This picture of causation, central to the physicalist approaches,
must be modified when it is carried over to the quantum regime.
There, the idea of a physical system following strict continuous
trajectories has long evaporated and in its place stands instead a
series of discrete points corresponding to specific measurement
events performed on the system. The inherent probabilistic nature
of the quantum world involving interference does not sanction any
definite interpolation between these points. Measurement is after
all a kind of interaction between the system and the measuring
apparatus that brings about an irreversible change to both.

Even though one may now find the notion of continuity dubious in
this domain, the concept of causal interaction survives seemingly
unscathed in the face of probabilistic indeterminism. In-between
measurements, the system is described by the continuous evolution
of the wavefunction as governed by the Schr\"{o}dinger equation.
However, this continuous evolution refers only to a distribution
of the different probabilities of obtaining various outcomes of
the measurements of a certain dynamical observable. It does not
represent an evolution of the successive dynamical values as
``possessed" by the system in time as in the classical case.
Quantum mechanically, the very act of measurements brings about
{\em irreversible} changes of the state of a physical system and
so interactions can be thought of as the cause as a result of
which the probability distributions of the values of dynamical
observables are altered. This is similar to the picture suggested
independently by Rudolf Haag\footnote{See, for example, Haag, R
(1990), ``Fundamental Irreversibility and the Concept of Events",
{\em Commun. Math. Phys.}, 132, p.245-251.} in a series of papers
which take the view that in the regime of low density, quantum
field theory describes a world where events are the collision
processes between particles and the particles themselves provide
the causal tie, i.e., the causal connection carrying the modified
dynamical structure.

\bigskip \bigskip
\hspace*{-5.5mm}{\bf V. Concluding Remarks}
\bigskip

Scientists seek causes for the purpose of  providing scientific
explanations. It is often regarded that effects follow necessarily
from causes. Indeed, it is the major contribution of Hume's theory
of causation to show that this is where we err and the inference
from causes to effects is not deductive but rather inductive in
nature. This immediately calls into question the idea of a
necessary connection which is thought to be the vital element of
causal relations. Russell went further and showed that there
clearly exist inconsistencies when we consider the causal relation
as one between events happening at discrete temporal intervals
superposed upon the temporal continuum.

Advocates of the class of the so-called physicalist approaches to
causation provide a promising framework for a theory of physical
causation in the face of Hume's and Russell's problems. The
problematic ``temporal gap" in Russell's analysis is closed by the
consideration of a continuum of events so that the temporal
continuum can be matched by the spatio-temporal continuous
character of physical processes. Causes, in these accounts, are
interactions which bring about irreversible changes from which a
causal order can be defined. This picture works well in the domain
of classical physics and requires appropriate modifications when
applied to the quantum regime.

Perhaps I should now leave the reader with the following thought.
The most startling case for causation is that of spontaneous decay
where an atom sits there for a while and then undergoes decay.
These are conceived to be ``uncaused" events by most philosophers
because the time when the atom is to decay cannot be known with
exactitude. Might the nucleus resemble Russell's ``static,
unchanging" event\footnote{It has been kindly pointed out to me by
Professor Morton Rubin that the atom is not really ``unchanging"
in Russell's sense: previous interactions with the vacuum must be
present in order to "prepare" the nucleus in an unstable state in
the first place. However, I would argue that this does not weaken
the thrust of Russell's argument, namely that even given that we
have a changing event, how might one explain how the atom comes to
decide at which exact moment in time it should undergo decay?},
sitting there for a period of time and suddenly undergoing decay?
Recalling that the major assumption that Russell makes is that
time is to be viewed as a mathematical continuum, it would be a
most interesting investigation to see whether adopting instead a
picture of time as discrete units will shed different light on
this problem. Once again, this illustrates how deeply the notion
of causation is connected with the very nature of space and time.

\end{document}